\documentclass[twocolumn]{article}  
\usepackage{graphicx}
\usepackage[T1]{fontenc}
\usepackage[subrefformat=parens]{subcaption}

\makeatletter
\let\cl@chapter\undefined
\makeatletter

\usepackage{amsmath,amssymb}   
\usepackage{amsthm} 
\usepackage{mathtools} 
\usepackage[numbers,sort&compress]{natbib}
\usepackage[margin=20mm]{geometry}
\usepackage{hyperref}
\hypersetup{colorlinks=true,pdfborder={0 0 0}}
\usepackage{siunitx}
\usepackage{cleveref}  
\usepackage{amsbsy}
\usepackage{bm}
\usepackage{url}
\usepackage{color}
\usepackage{multirow}
\usepackage{booktabs}
\usepackage{autobreak}
\allowdisplaybreaks 

\newcommand{\inlineTag}{
    \refstepcounter{equation}
    \bgroup\normalfont\normalcolor (\theequation)\egroup}

\crefname{equation}{Eq.}{Eqs.}%
\crefname{figure}{Fig.}{Figs.}%

\usepackage{algorithm}
\usepackage[noend]{algpseudocode}

\algnewcommand{\Break}{\textbf{break}}
\algrenewcommand\algorithmicindent{1em}

\providecommand{\argmin}{\operatornamewithlimits{argmin}} 

\everydisplay{\small} 

\newcommand\Erase{\bgroup\markoverwith{\textcolor{red}{\rule[.5ex]{2pt}{0.4pt}}}\ULon}

\begin{document}

    \title{
        Collision probability reduction method for tracking control in automatic docking / berthing using reinforcement learning
    }
    
    \author{
        Kouki Wakita $^{1}$\and
        Youhei Akimoto$^{2,3}$ \and
        Dimas M. Rachman $^{1}$ \and
        Yoshiki Miyauchi $^{1}$ \and
        Umeda Naoya $^{1}$\and
        Atsuo Maki $^{1}$
    }

    \date{%
    \flushleft{\footnotesize
        $^1$Osaka University, 2-1 Yamadaoka, Suita, Osaka, Japan \\%
        $^2$Faculty of Engineering, Information and Systems, University of Tsukuba, 1-1-1 Tennodai, Tsukuba, Ibaraki 305-8573, Japan \\
        $^3$RIKEN Center for Advanced Intelligence Project, 1-4-1 Nihonbashi, Chuo-ku, Tokyo 103-0027, Japan \\[2ex]%
        Email: kouki\_wakita@naoe.eng.osaka-u.ac.jp; maki@naoe.eng.osaka-u.ac.jp \\
    }}


    \maketitle

    \begin{abstract}
        Automation of berthing maneuvers in shipping is a pressing issue as the berthing maneuver is one of the most stressful tasks seafarers undertake. Berthing control problems are often tackled via tracking a predefined trajectory or path. Maintaining a  tracking error of zero under an uncertain environment is impossible; the tracking controller is nonetheless required to bring vessels close to desired berths. The tracking controller must prioritize the avoidance of tracking errors that may cause collisions with obstacles. This paper proposes a training method based on reinforcement learning for a trajectory tracking controller that reduces the probability of collisions with static obstacles. Via numerical simulations, we show that the proposed method reduces the probability of collisions during berthing maneuvers. Furthermore, this paper shows the tracking performance in a model experiment.            
    \end{abstract}
    
    \section{Introduction}\label{Sec-1}
        Autonomous vessels are becoming increasingly important due to crew shortages, cost reduction, and safety considerations. In particular, the automation and autonomy of berthing maneuvers are significant issues because berthing maneuvering is one of the most stressful maneuvers seafarers undertake.
        
        For a solution to the berthing control problem to be useful in practical situations, the solution must be generated in real time. However, this problem is challenging due to the computational constraints and the complexity and uncertainty of maneuverability. Therefore, the berthing control problem has often been tackled in two stages: \textit{trajectory planning} and \textit{trajectory tracking}. Here, in this paper, a \textit{trajectory} is defined as a time series of state variables, which are subject to spatial and temporal constraints that take into account the vessel dynamics. The generation of a trajectory is called \textit{trajectory planning,} and the tracking of a given trajectory is called \textit{trajectory tracking}. In particular, the desired trajectory of berthing maneuvering is called the \textit{berthing trajectory}. By preparing the berthing trajectory in advance, it is possible to solve the berthing control problem in real time only via tracking of the prepared trajectory.

        However, tracking the predefined berthing trajectory is not straightforward because the vessel speed changes significantly, the effect of wind disturbance increases at low-speed, and complex maneuvers, including backward and crabbing motion, are frequently required. Therefore, an essential task is to develop a tracking controller that can safely track a given berthing trajectory.
        
        \subsection{Related Research}\label{Sec-1.1}
            Considerable research has been conducted on trajectory tracking control of surface vessels \cite{Fossen2000,Sorensen2011,Ding2011,Zheng2014,Yang2014,Wen2019,Jiang2022,dimasSTABS}. For instance, there exists much research work on the application of nonlinear model predictive control to trajectory tracking control \cite{Zheng2014,Jiang2022}, and the development of tracking control based on backstepping control methods for robust control in situations where uncertainties due to disturbances exist \cite{Ding2011,Yang2014,Wen2019}. Studies have also been undertaken on dynamic positioning (DP) control, focusing on situations where the vessel speed is near zero \cite{Sorensen2011}.
            
            Research on adopting reinforcement learning (RL) to surface vessel control has recently increased \cite{Cheng2018,Martinsen2018a,Martinsen2018b,Martinsen2020RLtracking,Meyer2020a,Meyer2020b}. RL is an area of machine learning and can be used to find the optimal action maximizing a reward function in an environment containing uncertainty. Although RL requires training, RL does not require real time optimization computation because it uses function approximation via a neural network (NN). Martinsen et al. implemented an algorithm based on the deep deterministic policy gradient method to obtain a controller that minimizes the tracking errors in straight and curved paths in situations involving unknown currents \cite{Martinsen2018a,Martinsen2018b}. Subsequently, Martinsen et al. proposed an RL-based control scheme for trajectory tracking control that achieves both DP control and path following \cite{Martinsen2020RLtracking}. 
            
            However, berthing trajectories typically consist of complex motions such as turning, stopping the ship, backward motion, spinning, and crabbing motion. Therefore, training for trajectories that consist of any combination of motions is required.
    
            Additionally, in berthing maneuvers, the avoidance of collisions is also required. This is because a small tracking error can cause a collision in berthing maneuvers because the purpose of the berthing is to bring the vessel close to the desired berth, which in itself represents an obstacle. Moreover, it is impossible to keep tracking errors equal to zero in an uncertain environment. 
            
            Some studies on collision avoidance of surface vessels have already been undertaken \cite{Meyer2020a,Meyer2020b}. However, those studies treated path following and collision avoidance as dual-objective problem. In this case, path following and collision avoidance are in a trade-off relationship. The vessel may not be able to reach the berthing point due to avoiding the collision with the berth. Therefore, in berthing maneuvers, it is effective to avoid tracking errors that increase the probability of collisions.

        \subsection{Scope of this study}\label{Sec-1.2}
            The purpose of this study is to develop a methodology that avoids the tracking errors that may cause collisions with static obstacles. This paper proposes a training method for a trajectory tracking controller that reduces the probability of collision with static obstacles. The main contributions of this paper are as follows:
            \begin{enumerate}
               \item The generation of random trajectories generated from maneuvering simulation was introduced to permit trajectory tracking controllers to track complex trajectories.
               \item The development of a method to generate static pseudo-obstacles depending on the desired trajectory was proposed for training on the avoidance of collisions with obstacles.
               \item A measure representing the distance between a static obstacle and the vessel position was introduced to give the geometric information describing the obstacle to the controller.
               \item A new reward function was proposed to train the trajectory tracking controller to prioritize the avoidance of the variety of tracking errors that may cause collision.
               \item The potential of the proposed method to reduce the probability of collisions in berthing maneuvering was demonstrated via numerical simulations. Furthermore, the results of the model experiments conducted outdoors are shown.
            \end{enumerate}

            The proposed method requires a training environment including a maneuvering model and response characteristics of steering systems. However, it is unnecessary to prepare a large number of berthing trajectories and obstacle information for training; only berthing trajectory and obstacle information in the target port is required.
    
            The remainder of the paper is organized as follows: Section 2 describes the training method of the trajectory tracking controller. Section 3 describes how the trajectory tracking controller can be applied to berthing maneuvers. Section 4 reports and analyzes the numerical simulation and model experiment results of berthing maneuvering. Finally, Section 5 presents the conclusions of the study.

    \section{Method}\label{sec:tracking}
        \begin{figure}[t]
            \centering
            \includegraphics[width=0.95\hsize]{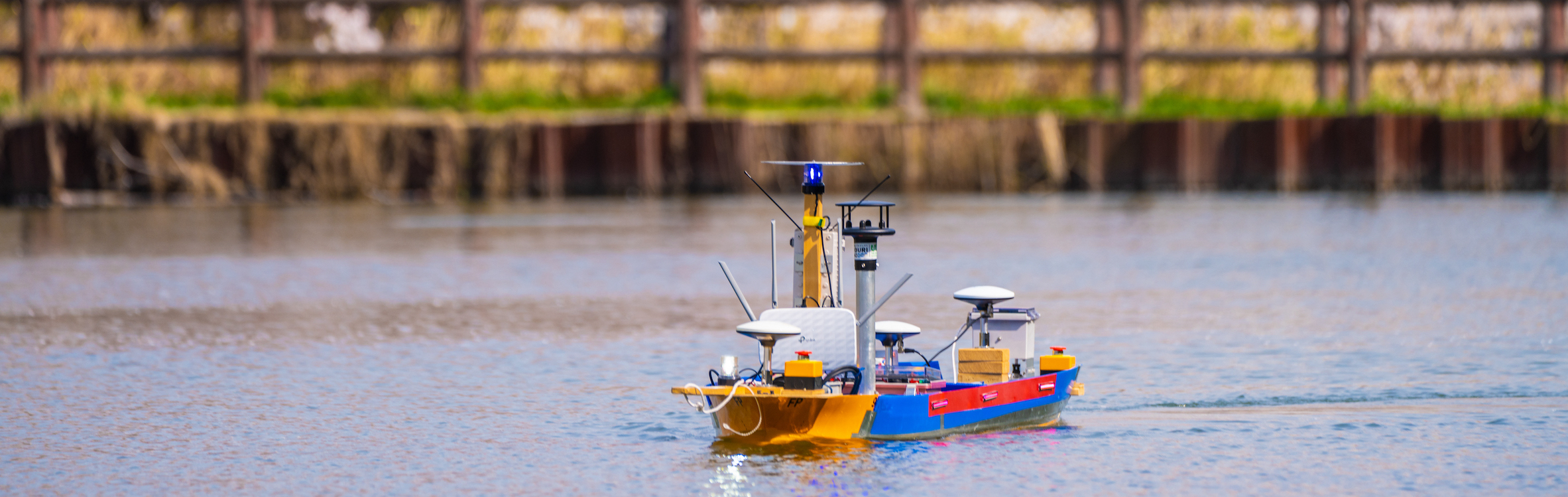}
            \caption{Subject ship in Inukai pond}
            \label{fig:subject_ship}
        \end{figure}
        
        \begin{table}[b]
            \centering
            \caption{Limits of the actuator state variables}
            \begin{tabular}{lll} \hline
                Item & Lower limit & Upper limit \\ \hline
                $\delta_{\mathrm{p}}$ & $-105 \ \mathrm{[deg.]}$ & $35 \ \mathrm{[deg.]}$ \\
                $\delta_{\mathrm{s}}$ & $-35 \ \mathrm{[deg.]}$ & $105 \ \mathrm{[deg.]}$ \\
                $n_{\mathrm{p}}$ & $10 \ \mathrm{[rps]}$ & $10 \ \mathrm{[rps]}$ \\
                $n_{\mathrm{bt}}$ & $-30 \ \mathrm{[rps]}$ & $30 \ \mathrm{[rps]}$ \\ \hline
            \end{tabular}
            \label{tab:controllimit}
        \end{table}
        
        The subject ship in this study is a model ship shown in \Cref{fig:subject_ship}. The length and breadth of this model ship are defined as $L$ and $B$, respectively. This ship is equipped with a single propeller, VecTwin rudders, and a bow thruster. The actuator states is defined as $\boldsymbol{u} \equiv \left( \delta_{\mathrm{p}}, \delta_{\mathrm{s}}, n_{\mathrm{p}}, n_{\mathrm{bt}} \right)^{\mathsf{T}} \in \mathbb{R}^4$,
        where $\delta_{\mathrm{p}}$ represents the rudder angle on the port side, $\delta_{\mathrm{s}}$ represents the rudder angle on the starboard side, $n_{\mathrm{p}}$ is the propeller revolution number, and $n_{\mathrm{bt}}$ is the bow thruster revolution number. In this study, the response characteristics of the steering rudder are taken into account. The control commands are defined as $\boldsymbol{u}_{\mathrm{cmd}} \equiv \left( \delta_{\mathrm{p},\mathrm{cmd}}, \delta_{\mathrm{s},\mathrm{cmd}}, n_{\mathrm{p},\mathrm{cmd}}, n_{\mathrm{bt},\mathrm{cmd}} \right)^{\mathsf{T}} \in \mathbb{R}^4$.
        The upper and lower limits of the actuator state variables are defined in \Cref{tab:controllimit}. Note that in this study, the propeller revolution number is constant at $10 \ \mathrm{rps}$ because the VecTwin rudder system enables various motions by changing the rudder angle while maintaining a constant propeller revolution number.
        
        \begin{figure}[t]
            \centering
            \includegraphics[width=0.65\hsize]{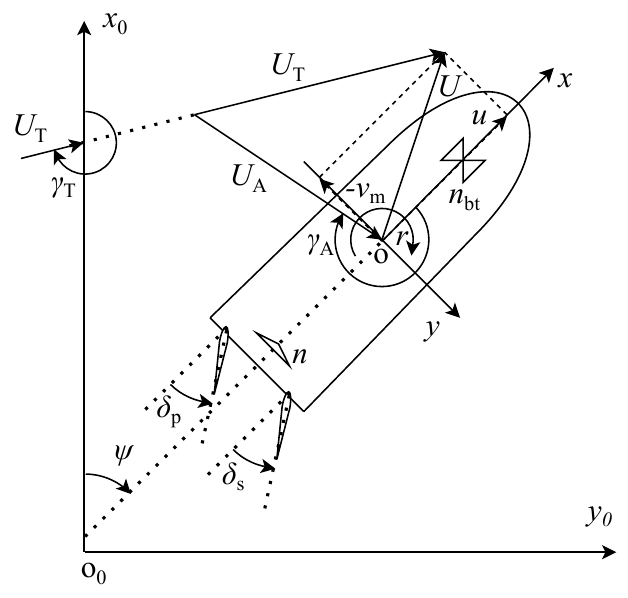}
            \caption{The coordinates systems used in this work.}
            \label{fig:coordinate_systems1p2r}
        \end{figure}
        
        In this study, the motion of the vessel in the harbor is modeled by surge--sway--yaw three-degrees-freedom equation of motion. The coordinate systems are earth-fixed coordinates, represented by $\mathrm{o}_{0}-x_{0}y_{0}$ and ship-fixed coordinates, represented by $\mathrm{o}-xy$, which has its origin on midship. The coordinate systems in this study is shown in \Cref{fig:coordinate_systems1p2r}. Here, the vessel state vector is defined as $\boldsymbol{x} \equiv \left( x_{0}, u, y_{0}, v_{\mathrm{m}}, \psi, r \right)^{\mathsf{T}} \in \mathbb{R}^6$,
        where $\psi$ is the heading angle, $u$ is the surge velocity, $v_{\mathrm{m}}$ is the sway velocity of the midship, and $r$ is the yaw angular velocity. For convenience, the vessel state vector $\boldsymbol{x}$ is divided into pose vector, $\boldsymbol{\eta} \equiv \left(x_{0}, y_{0}, \psi \right)^{\mathrm{T}} \in \mathbb{R}^{3}$, and the velocity vector, $\boldsymbol{v} \equiv \left(u, v_{\mathrm{m}}, r\right)^{\mathrm{T}} \in \mathbb{R}^{3}$. 
        The observed vessel state vector is defined as $\hat{\boldsymbol{x}} \in \mathbb{R}^6$. The vector of true wind speed and direction is defined as $\boldsymbol{w} \equiv\left(U_{\mathrm{T}}, \gamma_{\mathrm{T}}\right)^{\mathrm{T}} \in \mathbb{R}^{2}$, where $U_{\mathrm{T}}$ represents the true wind speed, and $\gamma_{\mathrm{T}}$ is the true wind direction. In this study, the true wind speed and direction are assumed to depend on time but not space. 
        
        \subsection{Optimization of the trajectory tracking controller}
            In this study, the trajectory tracking problem of minimizing the tracking error is formulated as a maximization problem related to a reward. This section describes the formulation used here.
            
            This paper focuses on the trajectory tracking of the pose vector, $\boldsymbol{\eta}$. Here, the desired pose vector is defined as $\boldsymbol{r} \in \mathbb{R}^{3}$. The time series of desired pose vector can be represented as follows:
            \begin{equation}
               \mathcal{R} = \left\{ \boldsymbol{r}_{k} \right\}_{k = 0, \ldots, N-1} \enspace ,
                \label{eq:desiredtrajectory}
            \end{equation}
            where $k$ denotes discrete-time steps, and $N$ is the total number of steps in the desired trajectory. The time of $k$-th step is defined as $t_{k}$, and the time step between $t_{k}$ and $t_{k+1}$ is defined as $\Delta t$. The time step between successive decisions in the RL algorithm is also equal to $\Delta t$.
            
            In this study, the geometric information related to static obstacles is assumed to be given. The static obstacles are represented by multiple polygons which can be defined,
            \begin{equation}
              \mathcal{O} = \bigcup_{k=1}^{N_{\mathrm{obs}}}{S_{i}} \enspace ,
                \label{eq:obstacle}
            \end{equation}
            where $S_{i}$ represents the $i$-th obstacle polygon, and $N_{\mathrm{obs}}$ represents the total number of obstacle polygons.
    
            In RL, the state variable for decision-making is often represented by $\boldsymbol{s}$ and the action variables by $\boldsymbol{a}$; these conventions is adopted in this work. In this study, the trajectory tracking controller makes decisions depending on the observed vessel state, $\hat{\boldsymbol{x}}$, the actuator state, $\boldsymbol{u}$, the desired trajectory, $\mathcal{R}$, and the static obstacle, $\mathcal{O}$. Thus, the state variable and the action variables at $k$-th step are defined as,
            \begin{equation}
                \left\{ \,
                    \begin{aligned}
                        \boldsymbol{s}_{k} &= \boldsymbol{f}_{s}\left(\hat{\boldsymbol{x}}_{k}, \boldsymbol{u}_{k}, \mathcal{R}, \mathcal{O}\right)\\
                        \boldsymbol{a}_{k} &= \boldsymbol{u}_{\mathrm{cmd},k} \enspace , \\
                    \end{aligned}
                \right.
                \label{eq:sa}
            \end{equation}
            where $\boldsymbol{f}_{s}$ is a function that generate the input of the controller; this function is described in \Cref{subsec:policy}. The action can then be determined as $\boldsymbol{a}_{k} = \boldsymbol{\mu}(\boldsymbol{s}_{k})$, 
            where $\boldsymbol{\mu}$ is the policy function. This function is the trajectory tracking controller, and is represented by a NN.
            
            This study considers trajectory tracking as an episodic task in RL. Thus, an objective function with a finite-time horizon can be defined as,
            \begin{equation}
                J(\boldsymbol{\mu}_{}) = E_{\tau_{} \sim \rho^{\boldsymbol{\mu}_{}}}\left[\sum^{N-1}_{k=0} \gamma^{k} r_{}\left(t_{k}, \boldsymbol{s}_{k}, \boldsymbol{a}_{k}\right) \right] \enspace ,
                \label{eq:evaluationfunction_obs}
            \end{equation}
            where $\gamma$ is the discount rate, and $r(\cdot)$ is the reward function, which depends on the tracking error, control effort, geometric relationship between vessel and obstacles, and the episode elapsed time, $t_{k}$. The reward function $r(\cdot)$ is described in detail in \Cref{subsec:reward}. $\tau_{}$ is a time series of the state and action variables throughout the episode and is defined as $\tau_{} = \left( \boldsymbol{s}_{0}, \boldsymbol{a}_{0}, \ldots, \boldsymbol{s}_{N-1}, \boldsymbol{a}_{N-1}, \boldsymbol{s}_{N}\right)$.
            $\rho^{\boldsymbol{\mu}_{}}$ can then be written as,
            \begin{equation}
                \rho^{\boldsymbol{\mu}} = p(\boldsymbol{s}_{0}) \prod^{T-1}_{k=0}p(\boldsymbol{s}_{k+1}|\boldsymbol{s}_{k},\boldsymbol{a}_{k}) \boldsymbol{\mu}\left(\boldsymbol{s}_{k}\right) \enspace ,
                \label{eq:probabilityepisode_obs}
            \end{equation}
            where $p(\boldsymbol{s}_{0})$ represents the probability distribution of the initial state, $\boldsymbol{s}_{0}$, observed from the environment; $p(\boldsymbol{s}_{k+1}|\boldsymbol{s}_{k},\boldsymbol{a}_{k})$ represents the state transition probability distribution. These distributions were implemented in the numerical simulations presented in this work and are described in detail in \Cref{subsec:env}. 
            The objective function described in \Cref{eq:evaluationfunction_obs} represents the expected cumulative reward obtained in an uncertain environment. In this study, the trajectory tracking controller that maximizes this objective function is calculated.
            Although solving this optimization problem is difficult, various suitable optimization methods have been proposed in the area of artificial intelligence. In this study, the twin-delayed deep deterministic policy gradient (\texttt{TD3}) algorithm \cite{fujimoto2018addressing} was used as the optimization method.
        
        \subsection{Training Environment}\label{subsec:env}
            This section describes the training environment used in this work. In training, the desired trajectory and the static obstacle are generated for each episode. After that, maneuvering simulations were conducted. The methods used for the generation of the desired trajectory and static obstacles are described in \Cref{subsubsec:desired_obsacle}. The environment of maneuvering simulation is described in detail in \Cref{subsubsec:stp}.
            
            \subsubsection{Desired trajectory and pseudo-obstacle} \label{subsubsec:desired_obsacle}
                In training, the desired trajectories, $\mathcal{R}$, are generated stochastically for each episode, and the static pseudo-obstacles, $\mathcal{O}$, are generated depending on the desired trajectories. This section describes the generation methods used for these variables. 
                
                The manner in which the desired trajectories were generated is described here. The target of this research is to train the trajectory tracking controller for berthing maneuvers. It is reasonable to train on a set of desired berthing trajectories in the target harbor. However, collecting many berthing trajectories is a time-consuming task. Moreover, training based on a small number of specific trajectories is undesirable because it may be necessary to change the desired trajectory due to practical considerations. Therefore, in this paper, trajectories obtained from a maneuvering simulation subject to random control input are introduced.
                
                The trajectories generated for training are required to include complex motions such as turning and backward motion. In this maneuvering simulation, control inputs were determined randomly at each time step of the simulation. The distribution followed by control inputs was determined so that the time average of the thrusts generated from actuators was around zero. Thus, the bow thruster revolution number was determined based on the uniform distribution over the interval listed in \Cref{tab:controllimit}. The port and starboard side rudder angles were determined based on the normal distributions of $\mathcal{N}(75, 30^2)$ and $\mathcal{N}(-75, 30^2)$, respectively. Here, the mean values of these distributions were set to take the value at which the net forces and the moment approached nearly zero.
                
                The initial state of the maneuvering simulation was determined randomly. In this study, the initial vessel velocity was determined based on the uniform distribution, whose interval is listed in \Cref{tab:initref}. The initial vessel pose was set to zero. In this maneuvering simulation, wind disturbance was neglected. 
                
                \begin{table}[h]
                    \centering
                    \caption{Initial velocity interval of the uniform distribution used to generate the desired trajectory.}
                    \begin{tabular}{lll} \hline
                        Item & Lower limit & Upper limit \\ \hline
                        $u$ & $-0.072 \ \mathrm{[m/s]}$ & $0.437 \ \mathrm{[m/s]}$ \\
                        $v_{\mathrm{m}}$ & $-0.07 \ \mathrm{[m/s]}$ & $0.07 \ \mathrm{[m/s]}$ \\
                        $r$ & $-0.1 \ \mathrm{[deg./s]}$ & $0.1 \ \mathrm{[deg./s]}$ \\ \hline
                    \end{tabular}
                    \label{tab:initref}
                \end{table}
                
                \begin{figure}[b]
                    \centering
                    \includegraphics[width=\hsize]{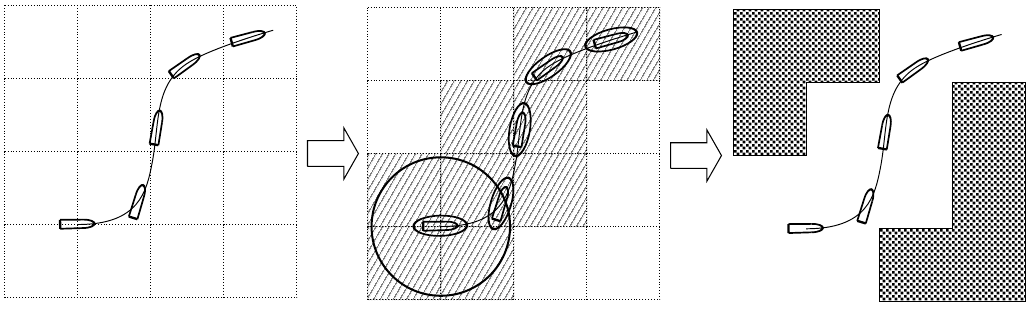}
                    \caption{The generation method of static pseudo-obstacles used in this work.}
                    \label{fig:pseudoobstacles}
                \end{figure}
                
                The generation method of the static obstacle is described. In this study, the static obstacles used in the training procedure were automatically generated according to the desired trajectory. Here, we impose that the desired trajectory does not contact any obstacle. We also defined that the generated obstacles consist of line segments that are all longer than the vessel length. A schematic of the proposed scheme is presented in \Cref{fig:pseudoobstacles} and can be summarized in the following steps:
                \begin{enumerate}
                    \item A grid is set based on earth-fixed coordinates $\mathrm{o}_{0}-x_{0}y_{0}$. Here, one of the grid points is set to coincide with the origin of the coordinate system used. The grid spacing of both the x-axis and y-axis is set to be $2L$, and the number of grid elements was set such that it covered the region where the desired trajectory exists.
                    \item The area where obstacles must not exist is then determined according to the desired trajectory. In this study, obstacles must not exist in the area through which the vessel passes when the motion of the desired trajectory is performed. Here, the vessel shape was represented by an ellipse whose semi-major axis was $0.75L$ and semi-minor axis was $B$. Additionally, obstacles must not exist in a circular area of radius $1.9L$ centered at the origin of the coordinate system; this condition avoids collisions due to initial tracking errors.
                    \item All grids through which the vessel does not pass are defined as static obstacles.
                \end{enumerate}
                
\begin{figure}[h]
                    \centering
                    \includegraphics[width=0.75\hsize]{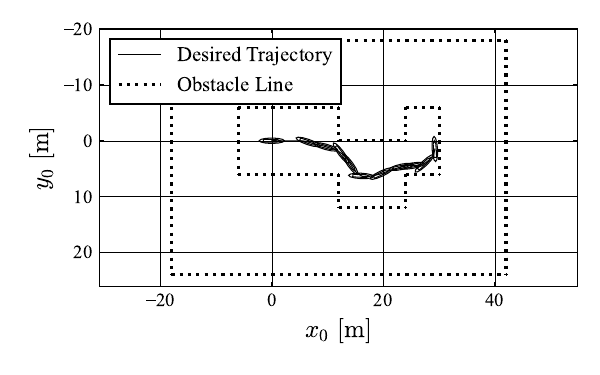}
                    \caption{An example of a desired trajectory with the static pseudo-obstacles used in training.}
                    \label{fig:pseudoobstacles_ex}
                \end{figure}

                An example of the generated trajectory and pseudo-obstacles is shown in \Cref{fig:pseudoobstacles_ex}.
                
            \subsubsection{Maneuvering simulation}\label{subsubsec:stp}
                The simulation environment was implemented considering a low-speed maneuvering model, steering response characteristics, wind disturbance, and artificial noise; the inclusion of all these elements led to a realistic environment. This section describes the models used to generate each of these elements in detail.
                
                First, the maneuvering model used in this study is described. The mathematical maneuvering group (MMG) model is widely used as a maneuvering model \cite{Ogawa1978}. In this study, a model based on the MMG model for low-speed maneuvering was used. The used MMG model consists of the following submodels: For the hull hydrodynamic forces, Yoshimura’s unified model \cite{Yoshimura2009} was used. To model the forces induced by the propeller, the model used in the work of Kang \cite{kang2008} was used.
                For the forces induced by the rudder, Kang's model \cite{kang2008}, which considers correlation effects between the port and starboard rudder, was applied. To model the forces induced by the bow thruster, Kobayashi's model \cite{Kobayashi1988}, which considers the forward speed effect, was applied. The force induced by the wind were modeled using Fujiwara's regression model \cite{Fujiwara1998}; this model estimates the wind force coefficients from the hull and superstructure geometry. All model parameters were derived based on captive model tests with the exception of the wind force coefficients. The MMG model used here is denoted as follows:
                \begin{equation}
                    \dot{\boldsymbol{x}} =  \boldsymbol{f}_{\mathrm{MMG}}\left(\boldsymbol{x}, \boldsymbol{u}, \boldsymbol{w}\right) \enspace ,
                    \label{eq:mmg}
                \end{equation}
                where the overdot, $\dot{ }$, denotes the derivative with respect to time, $t$. 
                
                Second, the response characteristics of actuator state, $\boldsymbol{u}$, are described. The rudder steering system with which the subject model ship was equipped has the characteristic of approaching a command value at a constant speed. Therefore, in this study, the response characteristics of rudder steering were modeled using the following expressions,
                \begin{equation}
                    \left\{ \,
                        \begin{aligned}
                            \dot{\delta}_{\mathrm{p}} &= K \cdot \text{sign}\left( \delta_{\mathrm{p},\mathrm{cmd}} - \delta_{\mathrm{p}} \right) \\
                            \dot{\delta}_{\mathrm{s}} &= K \cdot \text{sign}\left( \delta_{\mathrm{s},\mathrm{cmd}} - \delta_{\mathrm{s}} \right) \enspace , \\
                        \end{aligned}
                    \right.
                    \label{eq:rudderdelay}
                \end{equation}
                where $K$ is the rudder steering speed and is determined to be $20 \ \mathrm{deg./s}$. This value is taken based on measurements of the actual model ship system.
                The response characteristics of the bow thruster revolution numbers were neglected. The bow thruster revolution numbers are given as $n_{\mathrm{bt}} = n_{\mathrm{bt}, \mathrm{cmd}}$.

                The wind process was generated using the method proposed by Maki et al. \cite{maki2021wind}. In this method, one-dimensional filter equation using the wind speed spectrum approximated via Hino's spectrum \cite{hino1971spectrum} is numerically solved using Euler-Maruyama's method. This method can generate the wind process if the time-averaged true wind speed and direction are given. Here, the time-averaged true wind speed and direction are defined as $\bar{U}_{\mathrm{T}}$ and $\gamma_{\mathrm{T}}$, respectively. These variables are stochastically determined for each episode.
                
                In this study, the maneuvering model described in \Cref{eq:mmg} was solved numerically using the Runge-Kutta method, and the response characteristics described in \Cref{eq:rudderdelay} were solved analytically. Here, the time step of the numerical solutions is defined as $\Delta t_{\mathrm{sim}}$; this time step was set to be $0.1 \ \mathrm{s}$, which is shorter than the time step $\Delta t$. Here, the initial value of the vessel state is given such that the tracking error with the desired trajectory follows a uniform distribution over the interval listed in \Cref{tab:initenv}. The initial value of the actuator state is given as $(0.0, \ 0.0, \ 0.0, \ 0.0)$
                
                \begin{table}[t]
                    \centering
                    \caption{Uniform distribution intervals of the initial tracking error from the berthing trajectory.}
                    \begin{tabular}{lll} \hline
                        Item & Lower limit & Upper limit \\ \hline
                        $x_{0}$ & $-L \ \mathrm{[m]}$ & $L \ \mathrm{[m]}$ \\
                        $u$ & $-0.036 \ \mathrm{[m/s]}$ & $0.036 \ \mathrm{[m/s]}$ \\
                        $y_{0}$ & $-L \ \mathrm{[m]}$ & $L \ \mathrm{[m]}$ \\
                        $v_{\mathrm{m}}$ & $-0.007 \ \mathrm{[m/s]}$ & $0.007 \ \mathrm{[m/s]}$ \\
                        $\psi$ & $-10 \ \mathrm{[deg.]}$ & $10 \ \mathrm{[deg.]}$ \\
                        $r$ & $-0.1 \ \mathrm{[deg./s]}$ & $0.1 \ \mathrm{[deg./s]}$ \\ \hline
                    \end{tabular}
                    \label{tab:initenv}
                \end{table}
                
                Furthermore, process noise was added to the MMG model, and observation noise was added to the parameter describing the vessel state. The process noise and the observation noise were defined to follow a normal distribution, and their covariance matrices are represented by $\boldsymbol{\Sigma}_{\mathrm{sys}} \in \mathbb{R}^{6\times6}$ and $\boldsymbol{\Sigma}_{\mathrm{obs}} \in \mathbb{R}^{6\times6}$, respectively. The covariance matrix of the observed noise was determined by the nominal observation accuracy, which was described in the equipment specifications.
                
        \subsection{Input of the controller}\label{subsec:policy}
            This section describes the function $\boldsymbol{f}_{s}$ used in \Cref{eq:sa}. In this study, the trajectory tracking controller makes decisions depending on the vessel state, tracking error, and geometric relationship between the vessel and the obstacles. In the following, these elements determining the state, $\boldsymbol{s}$, are described.
            
            The tracking error is defined as the error between the vessel pose vector, $\boldsymbol{\eta}$, and the desired state, $\boldsymbol{r}$; this value is based on ship-fixed coordinate system $\mathrm{o}-xy$, and is represented as follows:
            \begin{equation}
                \boldsymbol{e}_{i,j}= 
                    \begin{bmatrix}
                        \cos \psi & \sin \psi & 0 \\
                        -\sin \psi & \cos \psi & 0 \\
                        0 & 0 & 1 \\
                    \end{bmatrix}
                \left(\boldsymbol{r}_{i}-\boldsymbol{\eta}_{j}\right) \enspace ,
                \label{eq:referror}
            \end{equation}
            where $i$ and $j$ represent the time step of the desired state and the vessel pose, respectively. The geometric relationship between vessel and obstacle is defined as the error between vessel positions and the obstacle point nearest to the desired positions; this parameter can be written as,
            \begin{equation}
                \tilde{\boldsymbol{e}}_{i,j}=
                    \begin{bmatrix}
                        \cos \psi & \sin \psi \\
                        -\sin \psi & \cos \psi \\
                    \end{bmatrix}
                \left(\boldsymbol{o}_{\mathrm{near},i}-\tilde{\boldsymbol{\eta}}_{j}\right) \enspace ,
                \label{eq:obserror}
            \end{equation}
            where $\tilde{\boldsymbol{\eta}}_{j}$ is defined as the position component of $\boldsymbol{\eta}_{j}$ and $\boldsymbol{o}_{\mathrm{near},i}$ is defined as,
            \begin{equation}
                \boldsymbol{o}_{\mathrm{near},i} = \argmin_{\boldsymbol{o} \in \mathcal{O}} \|\boldsymbol{o} - \tilde{\boldsymbol{r}}_{i}\| \enspace ,
                \label{eq:nearestobs}
            \end{equation}
            where $\tilde{\boldsymbol{r}}_{i}$ denotes the position component of $\boldsymbol{r}_{i}$.
            
            In this study, $\boldsymbol{s}_{k}$ is defined to consist of the tracking error, $\boldsymbol{e}_{i,j}$, the distance to the static obstacle, $\tilde{\boldsymbol{e}}_{i,j}$, the vessel velocity, $\boldsymbol{v}_{k}$ and the actuator state, $\boldsymbol{u}_{k}$. $\boldsymbol{s}_{k}$ is thus expressed as,
            \begin{equation}
                \begin{aligned}
                    \boldsymbol{s}_{k} = & \left( \boldsymbol{e}_{k,k}^{\mathsf{T}}, \ \boldsymbol{e}_{k+\frac{T_{1}}{\Delta t},k}^{\mathsf{T}}, \ \boldsymbol{e}_{k+\frac{T_{2}}{\Delta t},k}^{\mathsf{T}}, \ \cdots, \ \right. \\
                    & \left. \ \ \tilde{\boldsymbol{e}}_{k,k}^{\mathsf{T}}, \ \tilde{\boldsymbol{e}}_{k+\frac{T_{1}}{\Delta t},k}^{\mathsf{T}}, \ \tilde{\boldsymbol{e}}_{k+\frac{T_{2}}{\Delta t},k}^{\mathsf{T}}, \ \cdots, \ \boldsymbol{v}_{k}^{\mathsf{T}}, \ \boldsymbol{u}_{k}^{\mathsf{T}} \right)^{\mathsf{T}} \enspace ,
                \end{aligned}
                \label{eq:inputfeature}
            \end{equation}
            where $T_{1}, T_{2}, \cdots$ represent arbitrary time steps that are assumed to be positive multiples of $\Delta t$. Thus, this study also considers the error between the current vessel state and the future desired state as an the input of the controller. Therefore, the function $\boldsymbol{f}_{s}$ is defined via the use of \Cref{eq:referror,eq:obserror,eq:nearestobs,eq:inputfeature}

        \subsection{Reward function}\label{subsec:reward}
            This section describes the reward function used in this study. The reward function was designed to obtain a trajectory tracking controller that minimizes tracking errors and controls effort. Furthermore, we propose a method to preferentially minimize tracking errors that may lead to a collision with static obstacles.
            
            One measure of tracking error is the error norm of the vessel pose state, $\boldsymbol{\eta}$, and desired state, $\boldsymbol{r}$. However, this measure requires the time-consuming task of adjusting various weights due to the units being different for the information regarding position and heading angles. Therefore, in this study, the measures of the tracking error was defined based on the error of bow and stern positions that can be expressed as,
            \begin{equation}
                \left\{\begin{aligned}
                    e_{\mathrm{bow},k} &= \left\| \boldsymbol{g}_{\mathrm{bow}}(\boldsymbol{\eta}_{k}) - \boldsymbol{g}_{\mathrm{bow}}(\boldsymbol{r}_{k}) \right\|\\
                    e_{\mathrm{stern},k} &= \left\| \boldsymbol{g}_{\mathrm{stern}}(\boldsymbol{\eta}_{k}) - \boldsymbol{g}_{\mathrm{stern}}(\boldsymbol{r}_{k}) \right\| \enspace ,
                \end{aligned}\right.
                \label{eq:errorbowstern}
            \end{equation}
            where $\boldsymbol{g}_{\mathrm{bow}} \in \mathbb{R}^{2}$ and $\boldsymbol{g}_{\mathrm{stern}} \in \mathbb{R}^{2}$ denote the bow and stern positions, respectively; these variables are defined as,
            \begin{equation}
                \left\{\begin{aligned}
                    \boldsymbol{g}_{\mathrm{bow}}(\boldsymbol{\eta}) &= \left(x_{0}+\frac{1}{2}L\cos{\psi}, \ y_{0}+\frac{1}{2}L\sin{\psi}\right)^{\mathsf{T}} \\
                    \boldsymbol{g}_{\mathrm{stern}}(\boldsymbol{\eta}) &= \left(x_{0}-\frac{1}{2}L\cos{\psi}, \ y_{0}-\frac{1}{2}L\sin{\psi}\right)^{\mathsf{T}} \enspace .
                \end{aligned}\right.
                \label{eq:bowstern}
            \end{equation}
            
            \begin{figure}[t]
                \centering
                \includegraphics[width=0.5\hsize]{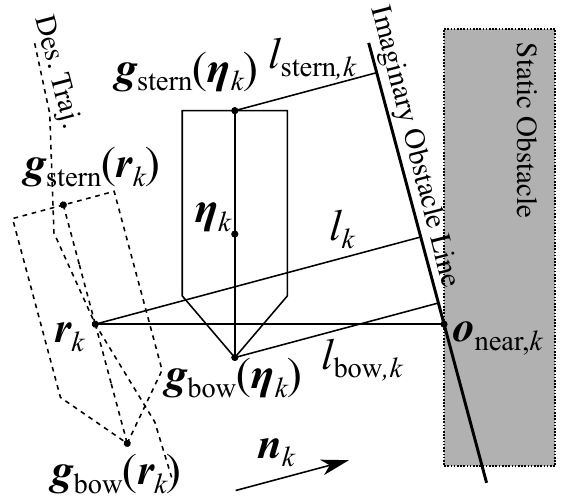}
                \caption{Illustration of the relationship between the vessel and static obstacle}
                \label{fig:col_reward}
            \end{figure}
            
            We also define the measure of the distance between the vessel and the static obstacles. Since the purpose of berthing is to get the vessel close to the berth, a measure that prevents the vessel from approaching the berthing point is undesirable. For this reason, a measure that has no trade-off between reducing tracking errors and avoiding the collision with the obstacles is defined here. 
            
            In this study, we introduced an imaginary obstacle line shown in \Cref{fig:col_reward}; this line passes through $\boldsymbol{o}_{\mathrm{near},k}$, and the angle of inclination of this line is equal to the desired heading angle. The normal vector on the $x_{0}y_{0}$ plane orthogonal to the imaginary obstacle line was also introduced. This normal vector is defined as $\boldsymbol{n}_{k} \in \mathbb{R}^2$, and can be calculated as according to,
            \begin{equation}
                \begin{bmatrix}
                \boldsymbol{n}_{k}\\0
                \end{bmatrix}
                = \begin{bmatrix}
                \boldsymbol{g}_{\mathrm{bow}}(\boldsymbol{r}_{k}) - \boldsymbol{g}_{\mathrm{stern}}(\boldsymbol{r}_{k})\\0
                \end{bmatrix} \times \boldsymbol{e}_{z} \enspace ,
                \label{eq:normalvec}
            \end{equation}
            where $\boldsymbol{e}_{z}$ is a unit vector orthogonal to the $x_{0}y_{0}$ plane, and this vector was $(0, 0, 1)^{\mathsf{T}}$.
            Here, the distance between the position of the desired state and the imaginary obstacle line was expressed as,
            \begin{equation}
                l_{k} = \left| \frac{\boldsymbol{n}_{k} \cdot \left(\tilde{\boldsymbol{r}}_{k} - \boldsymbol{o}_{\mathrm{near},k}\right)}{\left\|\boldsymbol{n}_{k}\right\|} \right| \enspace .
                \label{eq:lref}
            \end{equation}
            Besides, the distances from the actual bow and stern positions to the imaginary obstacle line were expressed as follows:
            \begin{equation}
                \left\{\begin{aligned}
                    l_{\mathrm{bow},k} &= \left| \frac{\boldsymbol{n}_{k} \cdot \left(\boldsymbol{g}_{\mathrm{bow}}(\boldsymbol{\eta}_{k}) - \boldsymbol{o}_{\mathrm{near},k}\right)}{\left\|\boldsymbol{n}_{k}\right\|} \right| \\
                    l_{\mathrm{stern},k} &= \left| \frac{\boldsymbol{n}_{k} \cdot \left(\boldsymbol{g}_{\mathrm{stern}}(\boldsymbol{\eta}_{k}) - \boldsymbol{o}_{\mathrm{near},k}\right)}{\left\|\boldsymbol{n}_{k}\right\|} \right| \enspace ,
                \end{aligned}\right.
                \label{eq:lbowstern}
            \end{equation}
            Then, the indicators that increase as the vessel approaches the obstacles compared to the desired position can be expressed as $l_{k} - l_{\mathrm{bow},k}$ and $l_{k} - l_{\mathrm{stern},k}$.
            
            Thus, the measures of nearness between the vessel and the static obstacles based on the desired position were defined as follows:
            \begin{equation}
                \left\{\begin{aligned}
                    c_{\mathrm{bow},k} &= \max \left\{0, \frac{l_{k} - l_{\mathrm{bow},k}}{l_{k}}\right\} \\
                    c_{\mathrm{stern},k} &= \max \left\{0, \frac{l_{k} - l_{\mathrm{stern},k}}{l_{k}}\right\} \enspace .
                \end{aligned}\right.
                \label{eq:cbowstern}
            \end{equation}
            Note that the cases where the vessel position is farther from the obstacles than the desired position were ignored since we focused on only reducing the tracking errors that increase the probability of collisions.
            
            Moreover, the tolerances of the measures described in \Cref{eq:errorbowstern,eq:cbowstern} were introduced. These tolerances are defined as,
            \begin{equation}
                \left\{\begin{aligned}
                    e_{\mathrm{tol},k} &= \left( e_{0} - e_{\infty} \right) \exp{\left( - b_{\mathrm{e}} t_{k} \right)} + e_{\infty} \\
                    c_{\mathrm{tol},k} &= \left( c_{0} - c_{\infty} \right) \exp{\left( - b_{\mathrm{c}} t_{k} \right)} + c_{\infty} \enspace ,
                \end{aligned}\right.
                \label{eq:tol}
            \end{equation}
            where $e_{0}$ and $c_{0}$ are the initial tolerances, $e_{\infty}$ and $c_{\infty}$ are the tolerances when the time of the episode elapses infinitely, and $b_{\mathrm{e}}$ and $b_{\mathrm{c}}$ are the attenuation factor of the tolerances.

            If the measures given in \Cref{eq:errorbowstern,eq:cbowstern} exceed these tolerances, no further rewards are given and the episode is stopped. In other words, the episode is terminated if one of the following inequalities is not satisfied:
            \begin{equation}
                \left\{\begin{aligned}
                    e_{\mathrm{bow},k} &< e_{\mathrm{tol},k} \\
                    e_{\mathrm{stern},k} &< e_{\mathrm{tol},k} \\
                    c_{\mathrm{bow},k} &< c_{\mathrm{tol},k} \\
                    c_{\mathrm{stern},k} &< c_{\mathrm{tol},k} \enspace .
                \end{aligned}\right.
                \label{eq:threshold}
            \end{equation}
            
            If a collision occurs, the episode is terminated. The collision detection area was considered to be equivalent to the area defined in \Cref{subsubsec:desired_obsacle}. If this area contacts an obstacle, the episode is terminated. 
            
            The reward function can thus be defined as follows:
            \begin{equation}
                \begin{aligned}
                    &r\left(t_{k}, \boldsymbol{s}_{k}, \boldsymbol{a}_{k}, \mathcal{R}, \mathcal{O}\right) = \frac{e_{\mathrm{tol},k} - e_{\mathrm{bow},k}}{e_{\mathrm{tol},k}} + \frac{e_{\mathrm{tol},k} - e_{\mathrm{stern},k}}{e_{\mathrm{tol},k}} + \\ &
                    \frac{c_{\mathrm{tol},k} - c_{\mathrm{bow},k}}{c_{\mathrm{tol},k}} + \frac{c_{\mathrm{tol},k} - c_{\mathrm{stern},k}}{c_{\mathrm{tol},k}} - \lambda \sum_{i=1}^{4} \left( \frac{u_{\mathrm{cmd},i} - u_{\mathrm{c},i}}{u_{\mathrm{std},i}} \right)^2 ,
                \end{aligned}
                \label{eq:reward_obs}
            \end{equation}
            where $\lambda$ is a positive constant value, $u_{\mathrm{c},i}$ represents the control input with the lowest control effort, and $u_{\mathrm{std},i}$ is a constant that normalizes each variable. Here, if $\lambda$ is large, the reward becomes negative. The controller may obtain more rewards when the episode is terminated early in such a case. For this reason, in this study, $\lambda$ is assumed to be small compared to the other terms.
    
    \section{Application to berthing maneuvers}
\begin{figure}[h]
            \centering
            \includegraphics[width=0.6\hsize]{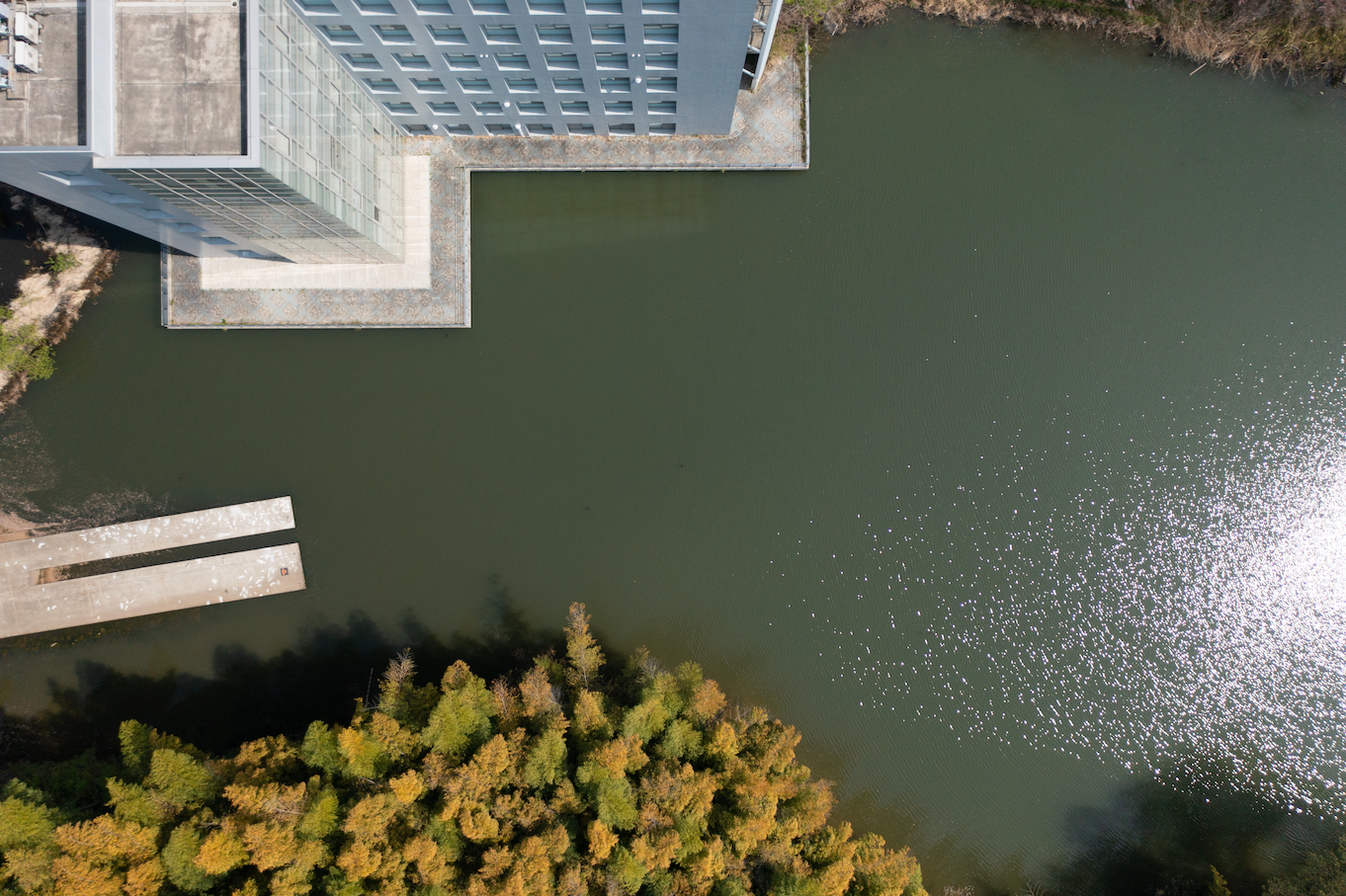}
            \caption{Port geometry of experimental pond}
            \label{fig:inukaipond}
        \end{figure}

        In this paper, the proposed methods were evaluated in terms of tracking the berthing trajectory. This section describes the generation method of the berthing trajectories and the development of tracking control methods for the berthing trajectories. The target port of this study is the Inukai pond, which exists at Osaka University. 
        This port geometry was shown in \Cref{fig:inukaipond}. 
        
        \subsection{Trajectory Planning}\label{subsec:planning}
            \begin{table}[t]
                \centering
                \caption{Initial and terminal conditions, as well as tolerances used in trajectory planning.}
                \begin{tabular}{cc}\hline
                    $\boldsymbol{\eta}_{\mathrm{init},i}$ & $\left( 24.0 \ \mathrm{[m]},i-5.0 \ \mathrm{[m]},\pi \ \mathrm{[rad]}\right), \enspace i=0,1,\cdots,10$ \\
                    $\boldsymbol{v}_{\mathrm{init}}$ & $\left( 0.291 \ \mathrm{[m/s]},0.0 \ \mathrm{[m/s]},0.0094 \ \mathrm{[rad/s]} \right)$ \\ \hline
                    $\boldsymbol{\eta}_{\mathrm{des},1}$ & $\left( -3.00 \ \mathrm{[m]},-0.7296 \ \mathrm{[m]},\pi \ \mathrm{[rad]}\right)$ \\
                    $\boldsymbol{\eta}_{\mathrm{des},2}$ & $\left( -3.00 \ \mathrm{[m]},-0.7296 \ \mathrm{[m]},0.0 \ \mathrm{[rad]}\right)$ \\
                    $\boldsymbol{\eta}_{\mathrm{des},3}$ & $\left( 0.7296 \ \mathrm{[m]},3.00 \ \mathrm{[m]},\frac{3\pi}{2} \ \mathrm{[rad]}\right)$ \\
                    $\boldsymbol{\eta}_{\mathrm{des},4}$ & $\left( 0.7296 \ \mathrm{[m]},3.00 \ \mathrm{[m]},\frac{\pi}{2} \ \mathrm{[rad]}\right)$ \\
                    $\boldsymbol{v}_{\mathrm{des}}$ & $\left( 0.0 \ \mathrm{[m/s]},0.0 \ \mathrm{[m/s]},0.0 \ \mathrm{[rad/s]} \right)$ \\ \hline
                    $\boldsymbol{\eta}_{\mathrm{tol}}$ & $\left( 0.02 \ \mathrm{[m]},0.02 \ \mathrm{[m]},0.0175 \ \mathrm{[rad]} \right)$ \\
                    $\boldsymbol{v}_{\mathrm{tol}}$ & $\left( 0.0141 \ \mathrm{[m/s]},0.0141 \ \mathrm{[m/s]},0.0094 \ \mathrm{[rad/s]} \right)$ \\ \hline
                \end{tabular}
                \label{tab:trajplanningsetting}
            \end{table}
            \begin{table}[t]
                \centering
                \caption{Limit of actuator state variables used in trajectory planning}
                \begin{tabular}{lll} \hline
                    Item & Lower limit & Upper limit \\ \hline
                    $\delta_{\mathrm{p}}$ & $-105 \ \mathrm{[deg.]}$ & $-45 \ \mathrm{[deg.]}$ \\
                    $\delta_{\mathrm{s}}$ & $45 \ \mathrm{[deg.]}$ & $105 \ \mathrm{[deg.]}$ \\
                    $n_{\mathrm{p}}$ & $10 \ \mathrm{[rps]}$ & $10 \ \mathrm{[rps]}$ \\
                    $n_{\mathrm{bt}}$ & $-15 \ \mathrm{[rps]}$ & $15 \ \mathrm{[rps]}$ \\ \hline
                \end{tabular}
                \label{tab:controllimittrajplan}
            \end{table}
            In this study, the desired trajectories for berthing maneuvers were generated according to the trajectory planning method proposed by Miyauchi et al. \cite{Miyauchi2022}. In this algorithm, the trajectory was explored in the framework of optimal control theory  \cite{Maki2020,Maki2020_2} with the use of covariance matrix adaptation evolution strategy (CMA-ES) (see, for example, Ref. \cite{Sakamoto2017}). 
            
            Here, the obtained trajectory is defined as,
            \begin{equation}
            \mathcal{R}^{\prime} = \left\{ \boldsymbol{r}^{\prime}_{i} \right\}_{i = 0, \ldots, N^{\prime}-1} \enspace ,
                \label{eq:desiredtrajectory_eval}
            \end{equation}
            where $N^{\prime}$ is the total number of time steps. Note that the time step of the desired trajectory obtained may differ from the decision-making time step $\Delta t$ defined in \Cref{eq:desiredtrajectory}. The time at which the i-th step occurs is defined as $t_{i}$, and the time step between $t_{i}$ and $t_{i+1}$ is defined as $\Delta t^{\prime}$.
            
\begin{figure}[h]
                \centering
                \includegraphics[width=0.8\hsize]{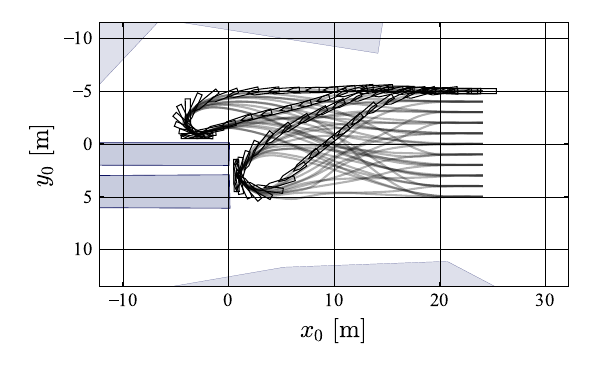}
                \caption{Desired trajectories generated by the trajectory planning procedure.}
                \label{fig:refpattern}
            \end{figure}

            We prepared 44 trajectories with 4 different terminal conditions and 11 different initial conditions; these conditions are listed in \Cref{tab:trajplanningsetting}. In the trajectory planning, the limits of actuator commands were changed from those listed in \Cref{tab:controllimit} to generate the trajectory with enough margins of control forces. The limit of the actuator states used is shown in \Cref{tab:controllimittrajplan}. This idea has been proposed by Kose et al. \cite{kose1986}. The obtained trajectories are shown in \Cref{fig:refpattern}. 
            
        \subsection{Selection of the desired state}
            In this study, we propose a method for the communication of the desired pose vector to the controller. In trajectory tracking, the desired state progresses depending on time to imitate the motion of the desired trajectory. However, this method may unnecessarily increase the vessel speed. In a berthing maneuver, the tracking control must not exceed the planned vessel speed. This is because the vessel may be unable to stop at the berthing point from a higher vessel speed. Therefore, the desired pose vector was chosen from the desired trajectory according to the vessel position: When the vessel pose is $\boldsymbol{\eta}_{k}$, the desired states are given by, 
            \begin{equation}
                    \left\{\begin{aligned}
                        \boldsymbol{r}_{k} &= \boldsymbol{r}^{\prime}_{i_{k}} \\
                        \boldsymbol{r}_{k + \frac{T_{1}}{\Delta t}} &= \boldsymbol{r}^{\prime}_{i_{k} + \frac{T_{1}}{\Delta t^{\prime}}} \\
                        \boldsymbol{r}_{k + \frac{T_{2}}{\Delta t}} &= \boldsymbol{r}^{\prime}_{i_{k} + \frac{T_{2}}{\Delta t^{\prime}}} \\
                        & \vdotswithin{=} \enspace , \\
                    \end{aligned}\right.
                    \label{eq:refik}
                \end{equation}
            where $i_{k}$ is defined as,
            \begin{equation}
                i_{k} = \argmin_{i_{k-1} < i < i_{k-1} + I} (\|\tilde{\boldsymbol{r}}^{\prime}_{i} - \tilde{\boldsymbol{\eta}}_{k}\|) \enspace ,
                \label{eq:ik}
            \end{equation}
            where $\tilde{\boldsymbol{r}}^{\prime}$ represents only the coordinate component of $\boldsymbol{r}^{\prime}$. $I$ represents the search range and should be defined to avoid shortcuts when desired trajectories intersect. Note that $T_{1}, T_{2}, \cdots$ must also be positive multiples of $\Delta t^{\prime}$.

        \subsection{Pseudo-obstacles in the target harbor}
            \begin{figure}[b]
                \centering
                \includegraphics[width=0.8\hsize]{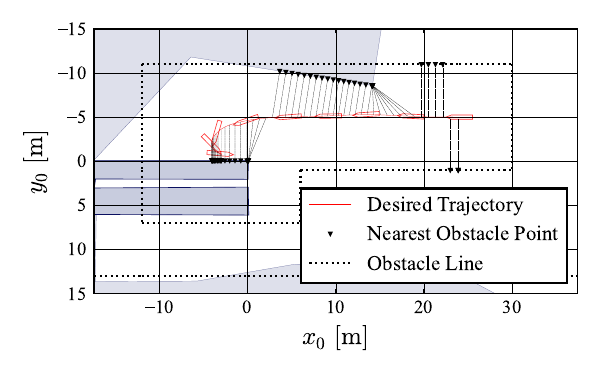}
                \caption{A sample of the generated pseudo-obstacles in the target harbor.}
                \label{fig:dummy}
            \end{figure}
            Tracking performance is degraded in situations that are different from those experienced in training due to the characteristics of NNs. The size of the space without the obstacles in the target port may differ from training. In this case, the scale of $\tilde{\boldsymbol{e}}_{k,k}, \tilde{\boldsymbol{e}}_{k+\frac{T_{1}}{\Delta t},k}, \tilde{\boldsymbol{e}}_{k+\frac{T_{2}}{\Delta t},k}, \cdots$ included in \Cref{eq:inputfeature} may be different within the training environment and the target port. Therefore, pseudo-obstacle was also generated in the target port to prevent tracking performance degradation.
            
            The method described in \Cref{subsubsec:desired_obsacle} would lead to the generation of pseudo-obstacles that would significantly affect the berthing maneuvering. Therefore, the ellipse area that was used to represent the vessel shape was changed to a circular area whose radius is $1.9L$. Since generated pseudo-obstacles are dummy, collisions with pseudo-obstacles are ignored. An example generated by the method described here is shown in \Cref{fig:dummy}.

    \section{Results}\label{sec:berthing}
        This section presents the training results of the proposed method and the results related to the berthing maneuver. We present the tracking results of the berthing trajectories obtained from both simulation and model experiments.
        
        In this study, for comparison of proposed methods, we prepared a trajectory tracking controller minimizing the tracking error and control effort in the absence of static obstacles. Here, $\tilde{\boldsymbol{e}}_{k,k}, \tilde{\boldsymbol{e}}_{k+\frac{T_{1}}{\Delta t},k}, \tilde{\boldsymbol{e}}_{k+\frac{T_{2}}{\Delta t},k}, \cdots$ included in \Cref{eq:inputfeature} are set to zero. The third and fourth terms of the proposed reward function described in \Cref{eq:reward_obs} is also set to zero. This trajectory tracking controller is referred to as Ctrl-w/o-OBST, whereas the trajectory tracking controller obtained by the proposed method was Ctrl-w/-OBST. 

        The training results of Ctrl-w/o-OBST and Ctrl-w/-OBST are shown in \Cref{sec:train}, and the tracking results of the berthing trajectories using Ctrl-w/o-OBST and Ctrl-w/-OBST are shown in \Cref{sec:berthingresult}.
        
        \subsection{Training results of trajectory tracking controller}\label{sec:train}
        
            \begin{table}[b]
                \centering
                \caption{Parameters describing the training environment.}
                \begin{tabular}{ll} \hline
                    $\bar{U}_{\mathrm{T}}$ & \multicolumn{1}{l}{given by weibull distribution whose} \\
                    & \multicolumn{1}{r}{shape and scale parameter are $2.0$ and $1.0$}\\
                    $\bar{\gamma}_{\mathrm{T}}$ & \multicolumn{1}{l}{given by uniform distribution} \\
                    & \multicolumn{1}{r}{whose interval is $[0, 2\pi)$}\\
                    $\boldsymbol{\Sigma}_{\mathrm{sys}}$ & $\mathrm{diag}\left( 0.0, 1.0\times10^{-4\cdot2}, \right.$ \\
                    & \multicolumn{1}{r}{$\left. 0.0, 1.0\times10^{-4\cdot2}, 0.0, 1.0\times10^{-3\cdot2} \right)$}\\
                    $\boldsymbol{\Sigma}_{\mathrm{obs}}$ & $\mathrm{diag}\left( 0.03^{2}, 0.02^{2}, 0.03^{2}, 0.02^{2}, 0.2^{2}, 0.2^{2} \right)$\\
                    $\Delta t$ & $5.0 \ \mathrm{[s]}$\\ \hline
                \end{tabular}
                \label{tab:envparam}
            \end{table}
            
            \begin{table}[b]
                \centering
                \caption{Numerical values used for the parameters present in \Cref{eq:inputfeature}.}
                \begin{tabular}{llll} \hline
                    $T_{1}$ & $5.0 \ \mathrm{[s]}$ & $T_{2}$ & $10.0 \ \mathrm{[s]}$ \\
                    $T_{3}$ & $20.0 \ \mathrm{[s]}$ & $T_{4}$ & $40.0 \ \mathrm{[s]}$ \\ \hline
                \end{tabular}
                \label{tab:ctrlparam}
            \end{table}

            This section shows the training result of the proposed method. Ctrl-w/-OBST and Ctrl-w/o-OBST were trained five times each. The training was undertaken for a total of $3.0\times10^7 \mathrm{s}$ of simulation time.
            
            The parameters used during training are summarized here. The environmental parameters are listed in \Cref{tab:envparam}, and the parameters describing the tracking controller and the NNs are listed in \Cref{tab:ctrlparam,tab:networkparam}, respectively. Here, the output dimension of the policy function is different from the dimension of the actuator command because the propeller revolution number is constant. The parameters of the objective function and the reward function are listed in \Cref{tab:jparam}. Here, the values of $e_{0}$ are set to be almost twice as large as the initial tracking error. The values of $b_{\mathrm{e}}$ and $b_{\mathrm{c}}$ are set such that the allowed range is halved after 50 seconds.
            
\begin{figure}[t]
                \centering
                \includegraphics[width=\hsize]{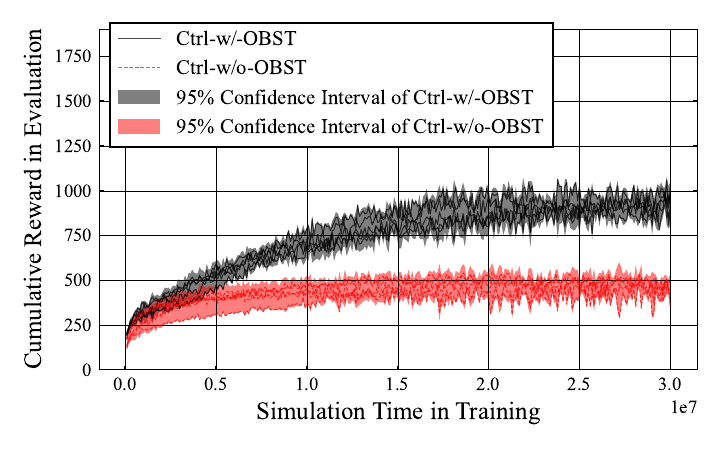}
                \caption{Average cumulative reward of the 20 episodes used for the evaluation.}
                \label{fig:cumreward}
            \end{figure}

            The parameters of the NNs acquired during the training were saved and then evaluated for all parameters of the NNs. In the evaluation, 20 episodes were simulated in the same environment as was used in the training episodes, and the average cumulative rewards were calculated. The average cumulative rewards and the 95\% confidence interval for the five training are shown in \Cref{fig:cumreward}. We see that the results do not deviate significantly from the $95\%$ confidence interval. Therefore, the parameter with the highest cumulative reward among the five training was selected as being the most appropriate method. Note that the cumulative rewards of Ctrl-w/-OBST and Ctrl-w/o-OBST cannot be compared directly as the reward functions are different.
            
            \begin{table}[t]
                    \small
                    \centering
                    \caption{Details of procedure used in the NNs. Values in bracket express the inputs into Ctrl-w/o-OBST.}
                    \begin{tabular}{c|ll|cc}\hline
                        & \multicolumn{2}{l|}{Units of layer} & \multicolumn{2}{|l}{Activation function} \\
                                & Policy & Q function & Policy & Q function \\ \hline
                        Input    & 32 (22) & 35 (25) & tanh & tanh  \\
                        Middle 1 & 256   & 256       & tanh & tanh  \\
                        Middle 2 & 256   & 256       & tanh & tanh  \\
                        Middle 3 & 256   & 256       & tanh & tanh  \\
                        output   & 3     & 1         & Sigmoid & Linear \\ \hline
                    \end{tabular}
                    \label{tab:networkparam}
                \end{table}
                
            \begin{table}[t]
                \centering
                \caption{Numerical values used for the parameters present in \Cref{eq:evaluationfunction_obs,eq:reward_obs}.}
                \begin{tabular}{m{1.5mm}m{12.0mm}m{1.5mm}m{12.0mm}m{2.0mm}m{6.5mm}m{3.5mm}m{6.5mm}} \hline
                    $\gamma$     & 0.99               & $\lambda$    & $1/300$ & $u_{\mathrm{c},1}$ & $-75$ & $u_{\mathrm{std},1}$ & $110$ \\
                    $e_{0}$      & $2L$ & $c_{0}$      & 1.0 & $u_{\mathrm{c},2}$ & $75$ & $u_{\mathrm{std},2}$ & $110$ \\
                    $e_{\infty}$ & $B/2$      & $c_{\infty}$ & 0.5 & $u_{\mathrm{c},3}$ & $10.0$ & $u_{\mathrm{std},3}$ & $1.0$ \\
                    $b_{\mathrm{e}}$      & $(\log2)/50$ & $b_{\mathrm{c}}$      & $(\log2)/50$ & $u_{\mathrm{c},4}$ & $0.0$ & $u_{\mathrm{std},4}$ & $30.0$ \\ \hline
                \end{tabular}
                \label{tab:jparam}
            \end{table}
            
            \begin{table*}[t]
                \small
                \centering
                \caption{The collision probability during  berthing for each wind speed and terminal conditions.}
                \begin{tabular}{l|cccccccc} \hline
                    \multirow{2}{*}{$\bar{U}_{\mathrm{T}}$} & \multicolumn{2}{c}{$\boldsymbol{\eta}_{\mathrm{des},1}$} & \multicolumn{2}{c}{$\boldsymbol{\eta}_{\mathrm{des},2}$} & \multicolumn{2}{c}{$\boldsymbol{\eta}_{\mathrm{des},3}$} & \multicolumn{2}{c}{$\boldsymbol{\eta}_{\mathrm{des},4}$} \\
                    & w/         & w/o        & w/                & w/o               & w/         & w/o        & w/         & w/o \\ \hline
                    0.0 $\mathrm{[m/s]}$   & 7.55 [\%]  & 94.73 [\%] & 38.55 (0.09) [\%] & 1.27 (0.00) [\%]    & 0.00 [\%]  & 96.36 [\%] & 0.27 [\%]  & 2.27 [\%] \\
                    0.5 $\mathrm{[m/s]}$   & 5.09 [\%]  & 90.73 [\%] & 38.27 (0.55) [\%] & 6.55 (0.00) [\%]    & 0.00 [\%]  & 93.09 [\%] & 0.64 [\%]  & 2.45 [\%] \\
                    1.0 $\mathrm{[m/s]}$   & 12.91 [\%] & 70.55 [\%] & 17.09 (15.82) [\%] & 29.18 (1.18) [\%]  & 0.45 [\%]  & 66.45 [\%] & 2.64 [\%]  & 5.64 [\%] \\
                    1.5 $\mathrm{[m/s]}$   & 38.64 [\%] & 63.27 [\%] & 49.82 (36.73) [\%] & 49.91 (33.00) [\%] & 36.82 [\%] & 65.73 [\%] & 37.09 [\%] & 51.09 [\%] \\ \hline
                \end{tabular}
                \label{tab:colprob}
            \end{table*}
        
        \subsection{Tracking results related to the berthing trajectories}\label{sec:berthingresult}
            This section shows the tracking results related to berthing trajectories. Here, a shorter decision-making interval for the evaluation was set to $\Delta t = 1.0 \ \mathrm{s}$, the time step of the berthing trajectory was set to $\Delta t^{\prime} = 0.2 \ \mathrm{s}$, and the search range used in \Cref{eq:ik} was set to $I = \Delta t / \Delta t^{\prime}$. 
            The simulation and model experiment results are shown in the following sections.
            
            \subsubsection{Simulation}
                \begin{figure}[t]
                    \centering
                    \includegraphics[width=0.9\hsize]{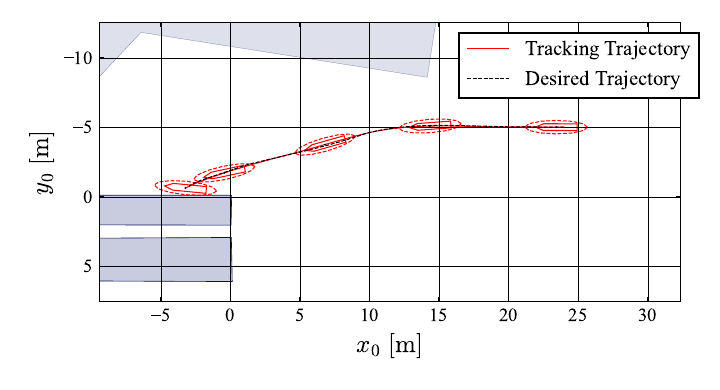}
                    \caption{An example of a result obtained from the simulation of a berthing maneuver using the Ctrl-w/o-OBST algorithm. The terminal condition of the desired trajectory is $\boldsymbol{\eta}_{\mathrm{des},1}$, and the average wind speed, $\bar{U}_{\mathrm{T}}$, is $0.0 \ \mathrm{m/s}$.}
                    \label{fig:sterncoll}
                \end{figure}
                \begin{figure}[t]
                    \centering
                    \includegraphics[width=0.9\hsize]{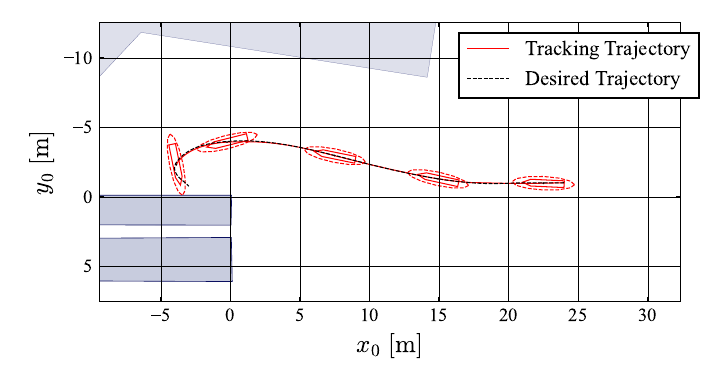}
                    \caption{An example of results obtained from the simulation of a berthing maneuver using the Ctrl-w/-OBST algorithm. The terminal condition of the desired trajectory is $\boldsymbol{\eta}_{\mathrm{des},2}$, and average wind speed, $\bar{U}_{\mathrm{T}}$, is $0.0 \ \mathrm{m/s}$.}
                    \label{fig:headcoll}
                \end{figure}
                \begin{figure}[t]
                    \centering
                    \includegraphics[width=0.9\hsize]{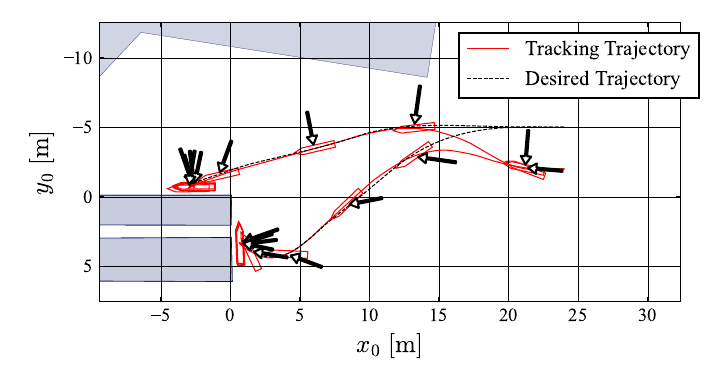}
                    \caption{An example of a result obtained from the simulation of a berthing maneuver using the Ctrl-w/-OBST algorithm. The terminal condition of the desired trajectory is $\boldsymbol{\eta}_{\mathrm{des},1}$ or $\boldsymbol{\eta}_{\mathrm{des},3}$, and average wind speed, $\bar{U}_{\mathrm{T}}$, is $1.0 \ \mathrm{m/s}$.}
                    \label{fig:result_case1}
                \end{figure}
                \begin{figure}[t]
                    \centering
                    \includegraphics[width=0.9\hsize]{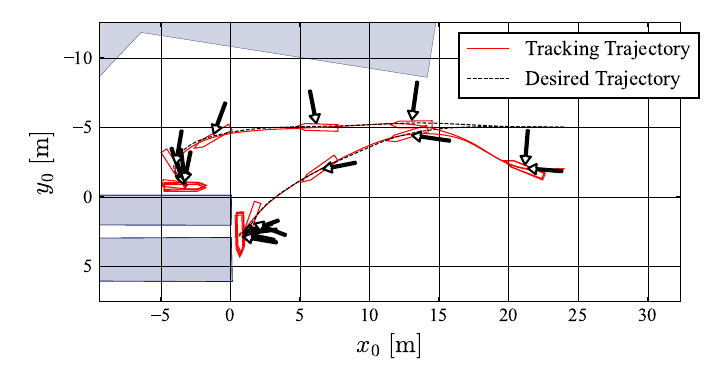}
                    \caption{An example of results obtained from the simulation of a berthing maneuver using the Ctrl-w/-OBST algorithm. The terminal condition of the desired trajectory is $\boldsymbol{\eta}_{\mathrm{des},2}$ or $\boldsymbol{\eta}_{\mathrm{des},4}$, and average wind speed, $\bar{U}_{\mathrm{T}}$, is $1.0 \ \mathrm{m/s}$.}
                    \label{fig:result_case2}
                \end{figure}

                We compare the results obtained using Ctrl-w/o-OBST and Ctrl-w/o-OBST in terms of collision probability during the berthing maneuvers to demonstrate the effectiveness of the method proposed here. The collision probability is defined here as the probability of colliding with a static obstacle before the end of a given simulation.
                
                In this study, 100 trials of berthing maneuvers were conducted to calculate the collision probability. The simulation was terminated if a collision occurred or the elapsed simulation time reached $250 \ \mathrm{s}$. The collision detection was conducted using the detection area defined in \Cref{subsubsec:desired_obsacle}. The initial value of the vessel state for each trial is given such that the tracking error with the given berthing trajectory follows a uniform distribution over the interval listed in Table 5. Collision probabilities were calculated for a given average wind speed to investigate the dependence of the collision probability on wind disturbance. Here, considering the wind pressure and the upper limit of the thrust that the subject ship can generate in bollard conditions, the collision probabilities were calculated with average wind speeds below $1.5 \ \mathrm{m/s}$. The obtained collision probabilities are listed in \Cref{tab:colprob}.
                
                \begin{figure}
                    \begin{minipage}[c]{\linewidth}
                        \centering
                        \includegraphics[width=\hsize]{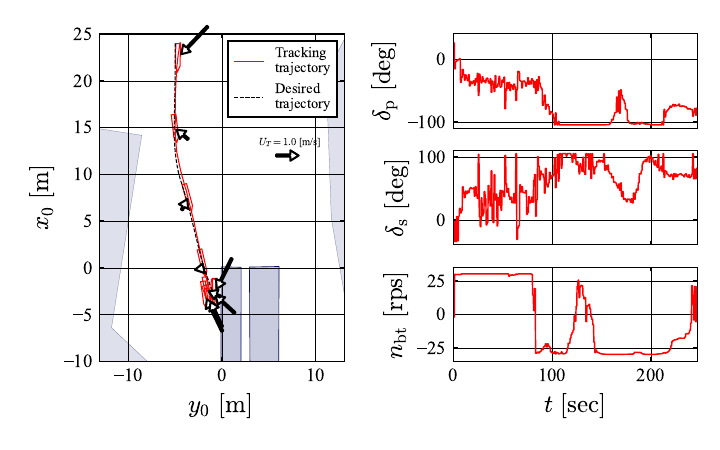}
                        \subcaption{Trajectories and the time series of control inputs}
                        \label{fig:exp_result_1_1}
                    \end{minipage}
                    \begin{minipage}[c]{\linewidth}
                        \centering
                        \includegraphics[width=\hsize]{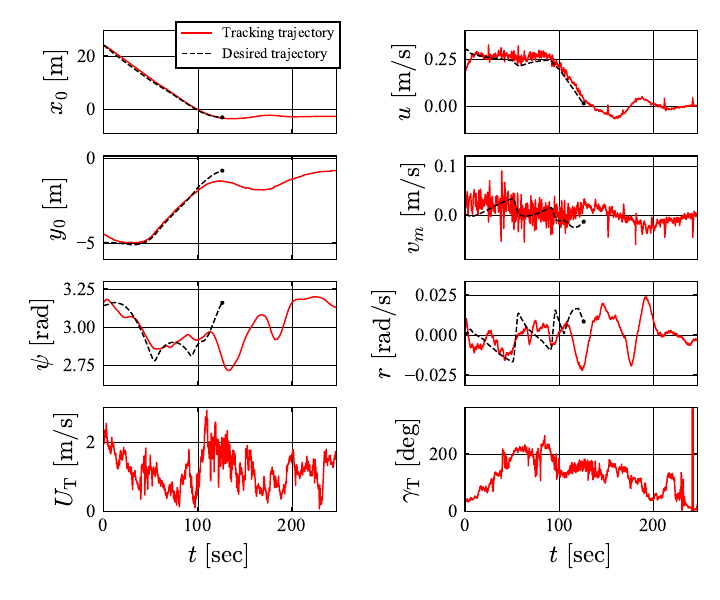}
                        \subcaption{The time series of the pose and velocity vector.}
                        \label{fig:exp_result_1_2}
                    \end{minipage}
                    \caption{The results of a model experiment showing the berthing maneuver obtained using the Ctrl-w/-OBST algorithm. The terminal condition of the desired trajectory was $\boldsymbol{\eta}_{\mathrm{des},1}$, and wind conditions were $\bar{U}_{\mathrm{T}}=1.19 \ \mathrm{[m/s]}, \bar{\gamma}_{\mathrm{T}}= 0.69\pi \ \mathrm{[rad]}$.}\label{fig:exp_result_1}
                \end{figure}
                
                \begin{figure}
                    \begin{minipage}[c]{\linewidth}
                        \centering
                        \includegraphics[width=\hsize]{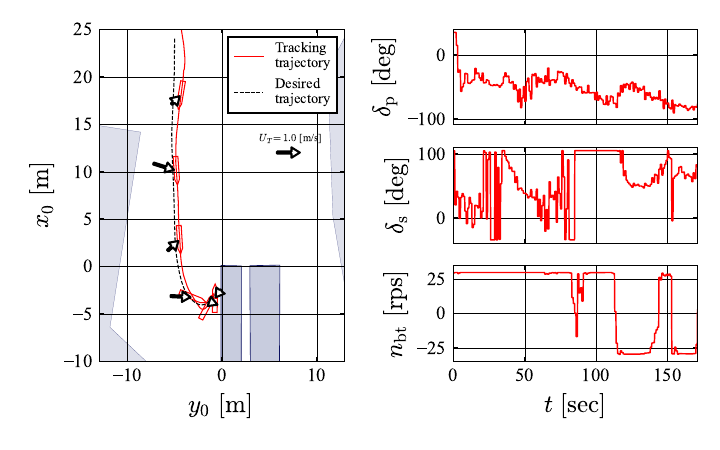}
                        \subcaption{Trajectories and the time series of control commands}
                        \label{fig:exp_result_2_1}
                    \end{minipage}
                    \begin{minipage}[c]{\linewidth}
                        \centering
                        \includegraphics[width=\hsize]{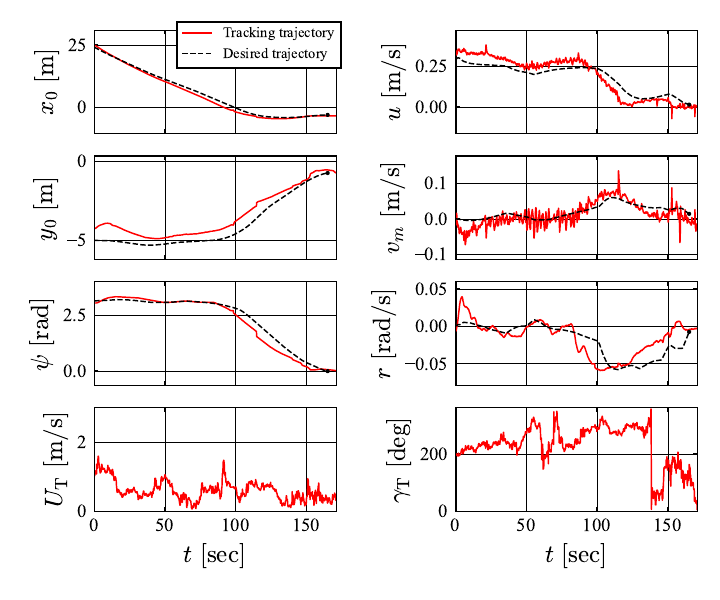}
                        \subcaption{The time series of the vessel pose and velocity.}
                        \label{fig:exp_result_2_2}
                    \end{minipage}
                    \caption{The results of a model experiment showing the berthing maneuver obtained using the Ctrl-w/-OBST algorithm. The terminal condition of the desired trajectory was $\boldsymbol{\eta}_{\mathrm{des},2}$, and wind conditions were $\bar{U}_{\mathrm{T}}=0.60 \ \mathrm{[m/s]}, \bar{\gamma}_{\mathrm{T}}=-0.62\pi \ \mathrm{[rad]}$.}\label{fig:exp_result_2}
                \end{figure}

                In the case where the Ctrl-w/o-OBST was used, we see that the collision probability is relatively high in the berthing trajectories whose terminal conditions are $\boldsymbol{\eta}_{\mathrm{des},1}$ or $\boldsymbol{\eta}_{\mathrm{des},3}$, whereas the collision probability is low for the berthing trajectories whose terminal conditions are $\boldsymbol{\eta}_{\mathrm{des},2}$ or $\boldsymbol{\eta}_{\mathrm{des},4}$. To investigate further the tracking performance of the Ctrl-w/o-OBST method, a tracking result without the initial tracking error is shown in \Cref{fig:sterncoll}. This result shows that the slight heading error causes a collision near the berthing point. Therefore, in berthing maneuvers, the collision probability increases due to small tracking errors even if the performance of the tracking controller is high.
                
                In the case where the Ctrl-w/-OBST was used, we see that the collision probability is relatively high in the berthing trajectories whose terminal conditions are $\boldsymbol{\eta}_{\mathrm{des},2}$. This is because, as shown in \Cref{fig:headcoll}, there are many cases in which the collision detection area on the bow side contacts the obstacle during turning. A slight tracking error causes a collision during turning because these berthing trajectories are generated considering the collision detection area described in \Cref{subsubsec:desired_obsacle}. The collision probabilities were therefore recalculated with a semi-major axis of the collision detection ellipse of $0.5L$. The obtained collision probabilities are listed in the brackets in \Cref{tab:colprob}. It was found under these conditions that the collision probability is similar to those obtained for the other terminal conditions.
                
                Comparing the results obtained using Ctrl-w/-OBST and Ctrl-w/o-OBST, we can see that the collision probabilities of Ctrl-w/-OBST are smaller than those obtained for Ctrl-w/o-OBST, except when the terminal condition is $\boldsymbol{\eta}_{\mathrm{des},2}$. In the case where the terminal condition is $\boldsymbol{\eta}_{\mathrm{des},2}$, the collision probability for Ctrl-w/-OBST is comparable to those obtained for other terminal conditions. This indicates that the proposed method preferentially avoids tracking errors that lead to collisions with obstacles, and that it is effective for berthing maneuvers.
                
                However, although the proposed method reduced the collision probability, it remains too high to be applicable to practical use. This study evaluated only whether a collision would occur during $250 \ \mathrm{s}$ and did not consider the effects of mooring time and fenders on the berth. Therefore, more detailed studies with a definition of successful berthing operations and the development of controllers that meet their safety requirements are needed.
                
                Finally, samples of the tracking results obtained using the Ctrl-w/-OBST are shown in \Cref{fig:result_case1,fig:result_case2}. In these samples, the initial error in the $y_{0}$-axis was $3.0 \ \mathrm{m}$, the average wind speed was $1.0 \mathrm{m/s}$, and the average wind direction was toward a berth at the berthing point. These results show that the controller was able to deal with an initial tracking error of about $L \ \mathrm{m}$.
                These results show that the same controller could track the desired trajectories with four types of terminal conditions.

            \subsubsection{Model experiment}
                In this study, the Ctrl-w/-OBST was evaluated via model experiments. In the model experiments, the berthing trajectories with terminal conditions of $\boldsymbol{\eta}_{\mathrm{des},1}$ and $\boldsymbol{\eta}_{\mathrm{des},2}$ were used. 
                The tracking results related to these berthing trajectories are shown in \Cref{fig:exp_result_1,fig:exp_result_2}, respectively.
                
                In the case of \Cref{fig:exp_result_1}, the wind speed increased as the model ship approached the berthing point, and reached $2.0 \ \mathrm{m/s}$. As a result, the wind disturbance caused the model ship to deviate from the desired trajectory, although the model ship reached the berthing point without collision. The probability of a collision would increase if the wind direction was different. Therefore, in this wind speed, a decision to stop the berthing maneuvers is required. Further research into the wind speed that permits safe berthing maneuvers is required. On the other hand, from the point of view of tracking control, this result shows that the controller is able to deal with tracking errors even during low-speed maneuvering.
                
                \Cref{fig:exp_result_2} shows the results of the model experiment conducted under relatively calm conditions. This result shows that the Ctrl-w/-OBST could track a berthing trajectory, including turning, and reduce the vessel speed to a value close to the allowable velocity, $\boldsymbol{v}_{\mathrm{tol}}$, near the berthing point.
                
    \section{Conclusion}\label{sec:conclusion}
        This paper proposed a training method based on RL for a trajectory tracking controller that reduces the probability of collision with static obstacles. This paper summarized the application of the obtained trajectory tracking controller to berthing maneuvers. To demonstrate the effectiveness of the proposed method, this paper showed the results of both simulations and model experiments related to the tracking of berthing trajectories in a target harbor. This work also showed that the proposed method reduces the probability of collision during in berthing maneuvers. Furthermore, this paper showed that the controller obtained by the proposed method was capable of following berthing trajectories and reducing the speed to a value close to the allowable velocity near the berthing point while avoiding collisions.

        In the proposed method, the controller was trained using obstacles consisting of line segments longer than the vessel length. However, actual harbors may have more complex shapes. Therefore, research related to training methods for obstacles consisting of shorter line segments is necessary. In this study, the controller was evaluated by considering the probability of collisions. However, collisions below a certain speed may be acceptable due to the presence of fenders. Therefore, further research on the definition of successful berthing is also necessary. Additionally, a decision to stop the berthing maneuvers is required for autonomous berthing when the effect of the wind disturbance is significant. Solving those problems is one of our future works.   
        
    \section*{Acknowledgements}
        This paper is a preprint published in the Journal of Marine Science and Technology, and the public version is available at (https://link.springer.com/article/10.1007/s00773-023-00962-5). 
        This study was conducted as part of Fully Autonomous Ship Program, ``MEGURI2040.'' This study was also conducted as collaborative research with Japan Hamworthy \& Co., Ltd. This study was supported by a Grant-in-Aid for Scientific Research from the Japan Society for the Promotion of Science (JSPS KAKENHI Grant \#19K04858 and \#22H01701). 
        The authors also would like to express gratitude to Enago (www.enago.jp) for reviewing the English language, and Mr. Satoru Konishi, Magellan Systems Japan Inc., for the technical support on GNSS measurement during the free run model test. Finally, the authors would like to thank Mr. Yuta Fueki, Mr. Nozomi Amano, and Mr. Hiroaki Koike, Osaka University, for supporting the free-run model test.



\begin{thebibliography}{10}

    \bibitem{Fossen2000}
    Thor~I. Fossen (2000)
    \newblock A survey on nonlinear ship control: from theory to practice.
    \newblock IFAC Proceedings Volumes, 33(21):1--16
    
    \bibitem{Sorensen2011}
    Asgeir~J. Sørensen (2011)
    \newblock A survey of dynamic positioning control systems.
    \newblock Annual Reviews in Control, 35(1):123--136
    
    \bibitem{Ding2011}
    Fuguang Ding, Yuanhui Wang, and Yong Wang (2011)
    \newblock Trajectory-tracking controller design of underactuated surface
      vessels.
    \newblock In: Proceedings of the OCEANS'11 MTS/IEEE KONA, pp 1--5
    
    \bibitem{Zheng2014}
    Huarong Zheng, Rudy~R. Negenborn, and Gabriel Lodewijks (2014)
    \newblock Trajectory tracking of autonomous vessels using model predictive
      control.
    \newblock IFAC Proceedings Volumes, 47(3):8812--8818
    
    \bibitem{Yang2014}
    Yang Yang, Jialu Du, Hongbo Liu, Chen Guo, and Ajith Abraham (2014)
    \newblock A trajectory tracking robust controller of surface vessels with
      disturbance uncertainties.
    \newblock IEEE Transactions on Control Systems Technology, 22(4):1511--1518
    
    \bibitem{Wen2019}
    Guoxing Wen, Shuzhi~Sam Ge, C.~L.~Philip Chen, Fangwen Tu, and Shengnan Wang (2014)
    \newblock Adaptive tracking control of surface vessel using optimized
      backstepping technique.
    \newblock IEEE Transactions on Cybernetics, 49(9):3420--3431
    
    \bibitem{Jiang2022}
    Xiaoyong Jiang, Langyue Huang, Mengle Peng, Zhongyi Li, and Ke-ji Yang (2022)
    \newblock Nonlinear model predictive control using symbolic computation on
      autonomous marine surface vehicle.
    \newblock Journal of Marine Science and Technology, 27(1):482--491
    
    \bibitem{dimasSTABS}
    Dimas~M. Rachman, Yoshiki Miyauchi, Naoya Umeda, and Atsuo Maki (2021)
    \newblock Feasibility study on the use of evolution strategy: {CMA-ES} for ship
      automatic docking problem.
    \newblock In: Proceedings of the 1st International Conference on the Stability and Safety of Ships and Ocean Vehicles.
    
    \bibitem{Cheng2018}
    Yin Cheng and Weidong Zhang (2018)
    \newblock Concise deep reinforcement learning obstacle avoidance for
      underactuated unmanned marine vessels.
    \newblock Neurocomputing, 272:63--73
    
    \bibitem{Martinsen2018a}
    Andreas~B. Martinsen and Anastasios~M. Lekkas (2018)
    \newblock Straight-path following for underactuated marine vessels using deep
      reinforcement learning.
    \newblock IFAC-PapersOnLine, 51(29):329--334
    
    \bibitem{Martinsen2018b}
    Andreas~B. Martinsen and Anastasios~M. Lekkas (2018)
    \newblock Curved path following with deep reinforcement learning: Results from
      three vessel models.
    \newblock In: Proceedings of the OCEANS 2018 MTS/IEEE Charleston, pp 1--8
    
    \bibitem{Martinsen2020RLtracking}
    Andreas~B. Martinsen, Anastasios~M. Lekkas, Sébastien Gros, Jon~Arne Glomsrud,
      and Tom~Arne Pedersen (2020)
    \newblock Reinforcement learning-based tracking control of usvs in varying
      operational conditions.
    \newblock Frontiers in Robotics and AI, vol 7.
    
    \bibitem{Meyer2020a}
    Eivind Meyer, Amalie Heiberg, Adil Rasheed, and Omer San (2020)
    \newblock Colreg-compliant collision avoidance for unmanned surface vehicle
      using deep reinforcement learning.
    \newblock IEEE Access, 8:165344--165364
    
    \bibitem{Meyer2020b}
    Eivind Meyer, Haakon Robinson, Adil Rasheed, and Omer San (2020)
    \newblock Taming an autonomous surface vehicle for path following and collision
      avoidance using deep reinforcement learning.
    \newblock IEEE Access, 8:41466--41481
    
    \bibitem{fujimoto2018addressing}
    S.~Fujimoto, H.~van Hoof, D.~Meger, Addressing function approximation error in actor-critic methods.
    \newblock In: Proceedings of the 35th International Conference on Machine Learning, Proceedings of Machine Learning Research, vol 80, pp 1587--1596
    
    \bibitem{Ogawa1978}
    A~Ogawa and H~Kasai (1978)
    \newblock On the mathematical model of manoeuvring motion of ships.
    \newblock International Shipbuilding Progress, 25(292):306--319
    
    \bibitem{Yoshimura2009}
    Yasuo Yoshimura, Ikao Nakao, and Atsushi Ishibashi (2009)
    \newblock Unified mathematical model for ocean and harbour manoeuvring.
    \newblock In: Proceedings of the International Conference on Marine Simulation and Ship Maneuverability, aug 2009, pp 116--124
    
    \bibitem{kang2008}
    Donghoon Kang, Vishwanath Nagarajan, Kazuhiko Hasegawa, and Masaaki Sano (2008)
    \newblock Mathematical model of single-propeller twin-rudder ship.
    \newblock Journal of marine science and technology, 13(3):207--222
    
    \bibitem{Kobayashi1988}
    Eiichi Kobayashi.
    \newblock A simulation study on ship manoeuvrability at low speeds (1988)
    \newblock Akishima Laboratory, Ocean Engineering Research Section, Mitsubishi
      Heave Industries Ltd. Published in: Mitsubishi Technical Bulletin No. 180
    
    \bibitem{Fujiwara1998}
    Toshifumi Fujiwara, Michio Ueno, and Tadashi Nimura (1998)
    \newblock Estimation of wind forces and moments acting on ships.
    \newblock Journal of the Society of Naval Architects of Japan, 1998(183):77--90
    
    \bibitem{maki2021wind}
    Atsuo Maki, Yuuki Maruyama, Leo Dostal, Sakai Masahiro, Sawada Ryohei, Sasa Kenji, and Naoya Umeda.
    \newblock Practical methodology of wind process generation for examining
      autonomous operating systems of ships.
    \newblock Under review.
    
    \bibitem{hino1971spectrum}
    M~Hino.
    \newblock Spectrum of gusty wind (1971)
    \newblock In: Proceedings of the 3rd International Conference on Wind
      Effects on Buildings and Structures, Tokyo, Japan, vol 77.
    
    \bibitem{Miyauchi2022}
    Yoshiki Miyauchi, Ryohei Sawada, Youhei Akimoto, Naoya Umeda, and Atsuo Maki (2022)
    \newblock Optimization on planning of trajectory and control of autonomous
      berthing and unberthing for the realistic port geometry.
    \newblock Ocean Engineering, 245:110390
    
    \bibitem{Maki2020}
    Atsuo Maki, Naoki Sakamoto, Youhei Akimoto, Hiroyuki Nishikawa, and Naoya Umeda (2020)
    \newblock Application of optimal control theory based on the evolution
      strategy (CMA-ES) to automatic berthing.
    \newblock Journal of Marine Science and Technology, 25(1):221--233
    
    \bibitem{Maki2020_2}
    Atsuo Maki, Youhei Akimoto, and Umeda Naoya (2021)
    \newblock Application of optimal control theory based on the evolution
      strategy (CMA-ES) to automatic berthing (part: 2).
    \newblock Journal of Marine Science and Technology, 26(3):835--845
    
    \bibitem{Sakamoto2017}
    Naoki Sakamoto and Youhei Akimoto (2017)
    \newblock Modified box constraint handling for the covariance matrix adaptation
      evolution strategy.
    \newblock In: Proceedings of the Genetic and Evolutionary Computation
      Conference Companion, Berlin, Germany, 15–19 July 2017, Companion material proceedings, pp 183--184

    \bibitem{kose1986}
    Kuniji Kose, Junji Fukudo, Kenji Sugano, Shigeru Akagi, and Mihoko Harada (1986)
    \newblock On a computer aided maneuvering system in harbours.
    \newblock Journal of the Society of Naval Architects of Japan, 1986(160):103--110
    
    \end{thebibliography}

\end{document}